# A Power and Area Efficient Lepton Hardware Encoder with Hash-based Memory Optimization

Xiao Yan, Zhixiong Di, Bowen Huang, Minjiang Li, Wenqiang Wang, Xiaoyang Zeng, Yibo Fan

***Abstract*—Although it has been surpassed by many subsequent coding standards, JPEG occupies a large share of the storage load of the current data hosting service. To reduce the storage costs, DropBox proposed a lossless secondary compression algorithm, Lepton, to further improve the compression rate of JPEG images. However, the bloated probability models defined by Lepton severely restrict its throughput and energy efficiency. To solve this problem, we construct an efficient access probability-based hash function for the probability models, and then propose a hardware-friendly memory optimization method by combining the proposed hash function and the N-way Set-Associative unit. After that, we design a highly parameterized hardware structure for the probability models and finally implement a power and area efficient Lepton hardware encoder. To the best of our knowledge, this is the first hardware implementation of Lepton. The synthesis result shows that the proposed hardware structure reduces the total area of the probability models by 70.97%. Compared with DropBox's software solution, the throughput and the energy efficiency of the proposed Lepton hardware encoder are increased by 55.25 and 4899 times respectively. In terms of manufacturing cost, the proposed Lepton hardware encoder is also significantly lower than the general-purpose CPU used by DropBox.***

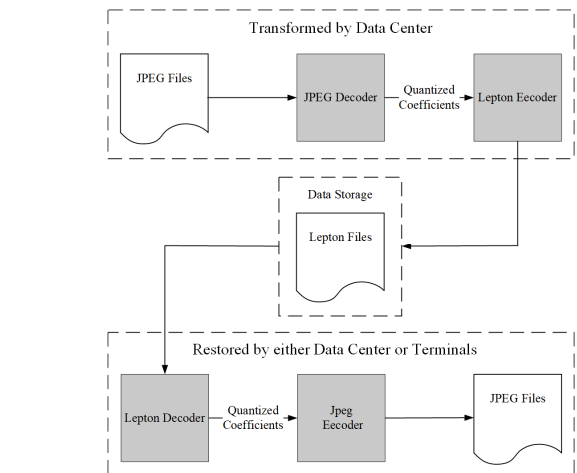

Fig. 1. Lepton-based efficient image storage solution. If the process of restoring Lepton files to JPEG files is done on the terminal, it can further save 23% of the transmission cost.

## I. INTRODUCTION

With the explosive growth of Internet applications, storage overhead has increasingly become the main cost of cloud storage, social networks, and e-commerce platforms. Although there are already a large number of new image compression standards, such as JPEG2000 [1], BPG [2], and WebP [3], which have proven to be far superior to JPEG's compression performance, JPEG [4] is still widely adopted due to simplicity and royalty-free. In terms of cloud storage services, according to Dropbox's statistics, JPEG images account for approximately 35% of the amount of data stored by individual users [5]. For e-commerce platforms like Alibaba, the proportion of JPEG images in the entire data storage is even higher. To store images more efficiently, DropBox proposed a lossless encoding algorithm, Lepton, to perform a secondary compression on JPEG images. Diversified test results show that Lepton can compress JPEG images by nearly 23% [5].

The transformation, storage, and recovery processes of the image files are shown in Fig. 1. The operation on the data center side is mainly composed of JPEG decoding and Lepton encoding. The data center uses a standard JPEG decoder to decode the JPEG file into 8x8 quantized coefficient blocks and then uses the Lepton encoding algorithm to re-encode these quantized coefficients into a more compact Lepton file. The backend servers only need to save these Lepton files, which significantly reduces the storage cost. When the user requests

This work was supported in part by Alibaba Innovative Research (AIR) Program, in part by the Shanghai Science and Technology Committee (STCSM) under Grant 19511104300, in part by National Natural Science Foundation of China under Grant 61674041, in part by the Innovation Program of Shanghai Municipal Education Commission, in part by the Fudan University-CIOMP Joint Fund (FC2019-001). (Corresponding author: Yibo Fan)

Xiao Yan, Yibo Fan, Bowen Huang, Minjiang Li, and Xiaoyang Zeng are with the State Key Laboratory of ASIC and System, Fudan University, Shanghai 200433, China (e-mail: {19112020084, fanyibo, 19212020120, 17212020017, xyzeng}@fudan.edu.cn;).
Zhixiong Di is with School of Information Science and Technology, Southwest Jiaotong University, Chengdu, China (e-mail: dizhixiong2@126.com)
Wenqiang Wang is with Alibaba Group, Hangzhou, Zhejiang 311121, China (e-mail: channing.wwq@alibaba-inc.com)

image data, DropBox decodes the Lepton files into quantized DCT coefficients and then encodes these coefficients into the original JPEG file. The task of restoring the Lepton file to a JPEG image can be executed either on the server-side or by the terminal. Performing restoration on the terminal side can further save transmission bandwidth.

To obtain significant compression gains, Lepton employs the binary arithmetic coding algorithm and greatly expands the probability models, resulting in a huge cost in storage and calculation. Binary arithmetic coding is a high-efficiency entropy coding algorithm, which needs to estimate the probability distribution of the input data [6]. Lepton defines a bunch of huge probability models, containing a total of 721564 bins to track the spatial correlation of the image. During the arithmetic coding, the probability models are used to record the history data and estimate the probability of 0 and 1 in the next bit.

In DropBox's storage system, Lepton is executed by the file servers or a dedicated server cluster, sharing production hosts with other memory-hungry processes. Therefore, the memory consumption caused by the probability models severely restricts the execution efficiency of Lepton. According to Horn et al. [5], an Intel Xeon E5 2650 v2 at 2.6GHz can only process 5.75 pictures per second and one kWh can be traded for an average of 72,300 Lepton conversions of images sized at an average of 1.5 MB each. It can be seen that the limited throughput and energy efficiency are the main bottleneck of Lepton application.

In this work, we propose an optimized Lepton hardware encoder to significantly increase the throughput of Lepton encoding while greatly reducing its energy consumption. Similar to the software solution, the huge probability models also pose severe challenges to the hardware encoder. Therefore, the memory optimization method of the probability models becomes a key issue. Although the hierarchical memory architecture is usually used in hardware to implement high-efficient memory access [7], designing a dedicated hierarchical memory architecture for the Lepton hardware encoder will cause an unacceptable area overhead. For software implementations, hash mapping is usually used to map a large range of indexes to a compact physical memory space [8] [9] [10], but it is usually difficult to construct a nearly collision-free and hardware-friendly hash function. By observing the Lepton coding process, we find that there are significant differences between the access probabilities of different bins in the probability models. Based on the statistical characteristics of the probability models accessing, we propose a memory optimization method for the probability models by combining hash mapping and N-way Set-Associative unit, and then design a high throughput, high energy efficiency, and low-cost Lepton hardware encoder.

The main contributions of this work are: 1. we propose an efficient hash function for Lepton's probability models. Kraska et al. [11] pointed out that for read-only databases, the cumulative distribution function (CDF) can be used as a hash function for data index. Inspired by this, this work further constructs a high-efficient hash function for write-heavy sparse data access. 2. By combining the hash function with the N-way

Fig. 2.  Regions of 8x8 quantized coefficient block. Lepton encodes the quantized coefficients sequentially according to their region, and the number in this figure is the coding order within each region.

Set-Associative Unit, we propose a hardware-friendly memory optimization method and reduce the area of the probability models by 70.97%. We argue that the proposed memory optimization method is also meaningful for software solutions and can benefit a large number of sparse memory access in specific domains. 3. Based on the memory optimization method, we design a low-cost Lepton hardware encoder, which greatly improves the throughput and energy efficiency of Lepton coding. Compared with DropBox's software solution, the throughput of the Lepton hardware encoder is increased by more than 55.25 times, and the power consumption is reduced by 4899 times. Besides, the cost of Lepton hardware encoder is much lower than the general-purpose CPU used by DropBox.

The rest of this paper is organized as follows. Section II introduces the basics of Lepton coding and Section III reviews the related works of memory management. The proposed hash function and memory optimization method is described in detail in Section IV. Section V introduces the VLSI architecture of the Lepton hardware encoder. Experiments and results are presented in Section VI. Finally, conclusions are drawn in Section VII.

## II. BASICS OF LEPTON CODING

### A. The main steps of Lepton coding

As shown in Fig. 2, the 8x8 quantized coefficient block obtained by JPEG decoding is divided into DC, x edge, y edge, and 7x7 AC regions, and then Lepton losslessly re-encodes these coefficients through binarization and binary arithmetic coding to obtain a higher compression rate.

The syntax elements and coding order of Lepton are shown in Table I. The 8x8 blocks are coded sequentially in the raster order. Lepton predicts the DC coefficient based on the neighbor blocks and then encodes its residual value. For x edge, y edge, and 7x7 AC regions, Lepton starts from the first coefficient and encodes to the last non-zero coefficient. Once there is no non-zero coefficient in a region, Lepton will skip the coefficient encoding process.

Before binary arithmetic coding, the data needs to be binarized. For the number of non-zero coefficients, Lepton directly uses its fixed-length binary code as the result of binarization. For the coefficient data, Lepton converts it into exponent, sign, and residual by using the Exp-Golomb code[12].

### B. Binary arithmetic coding and probability models

The crucial improvement of Lepton is to replace the Huffman coding in JPEG with the binary arithmetic coding. The binary arithmetic coding is the most widely used entropy coding method in image compression, its coding performance depends largely on the accuracy of the probability model. The probability model consists of a large number of bins, and each bin records the corresponding historical information in the encoding process. We take the exponent of the coefficients in 7x7 AC region, namely exp_7x7_data, as an example to illustrate Lepton's probability models. All bits of exp_7x7_data use the same probability model index

$$Index_{orig} = \{flag_c, prior, num\_nz\_left, idx_{zigzag}\}. \quad (1)$$

where $flag_c$ is the flag of the color component, $prior$ is calculated from the reference blocks, $num\_nz\_left$ is determined by the number of the left non-zero coefficients in the 7x7 AC region, $idx_{zigzag}$ is the zigzag index of the current coefficient. Table II shows the relationship between the probability models and each bit when exp_7x7_data is 0x3ff, 0x3e, and 0x6. The most significant bit of exp_7x7_data corresponds to probability model exp_7x7_0, and the subsequent bits are arranged in sequence. It can be seen that exp_7x7_10 will only be used when exp_7x7_data is large enough while exp_7x7_0 is always used. Denote the bit width of $Index_{orig}$ as $bw$, then directly using $Index_{orig}$ will require each probability model to have $2^{bw}$ bins. Since each field of $Index_{orig}$ has a limited range, we use

$$Index_{new} = \sum_{i=0}^{n}(Filed_i * \prod_{j=i+1}^{n} Range_j) \quad (2)$$

where $Filed_i$ is the value of the i-th field of the $Index_{orig}$, $Range_j$ is the number of valid values for the j-th field, to make the probability models more compact.

TABLE I
THE SYNTAX ELEMENTS AND CODING ORDER OF LEPTON

| Coding Order | Syntax Elements | Description |
| --- | --- | --- |
| 1 | num_nonzero_7x7 | number of non-zero coefficients in 7x7 AC region |
| 2 | 7x7 AC coefficients | from the first to the last non-zero coefficient in 7x7 AC region |
| 3 | num_nonzero_x_edge | number of non-zero coefficients in x edge region |
| 4 | x edge coefficients | from the first to the last non-zero coefficient in x edge region |
| 5 | num_nonzero_y_edge | number of non-zero coefficients in y edge region |
| 6 | y edge coefficients | from the first to the last non-zero coefficient in y edge region |
| 7 | DC residual | the residual value of DC |

TABLE II
CORRESPONDENCE BETWEEN EXP_7X7_DATA AND ITS PROBABILITY MODELS

| probability models | The value of exp_7x7_data | | |
| --- | --- | --- | --- |
| | 0x3ff | 0x3e | 0x6 |
| exp_7x7_0 | 1 | 1 | 1 |
| exp_7x7_1 | 1 | 1 | 1 |
| exp_7x7_2 | 1 | 1 | 0 |
| exp_7x7_3 | 1 | 1 | NA |
| exp_7x7_4 | 1 | 1 | NA |
| exp_7x7_5 | 1 | 0 | NA |
| exp_7x7_6 | 1 | NA | NA |
| exp_7x7_7 | 1 | NA | NA |
| exp_7x7_8 | 1 | NA | NA |
| exp_7x7_9 | 1 | NA | NA |
| exp_7x7_10 | 1 | NA | NA |

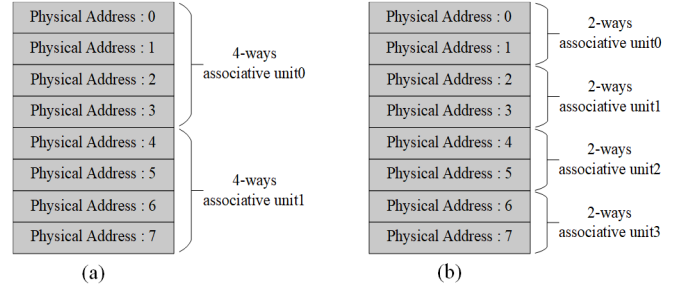

Fig. 3. Structure of typical N-way Set-Associative units. Suppose we need to map the input index to 8 physical addresses. (a) shows the case of two 4-way Set-Associative units. For the input index, we first determine its corresponding 4-ways Set-Associative unit, and then the index can use any memory space in this unit. Similarly, (b) shows the case of four 2-ways Set-Associative units.

To obtain significant compression gains, Lepton uses a bunch of probability models to independently record the probability distribution for each bit of each syntax element. The probability models defined by Lepton [13] are shown in Table III. After the conversion of (2), Lepton's probability models contain 685034 bins and require a total of more than 11M bits memories. It will be an unacceptable cost for hardware design.

To solve this problem, we establish an efficient hash function and combine the N-way Set-Associative unit to propose a hardware-friendly memory optimization method. After that, we design a highly reusable structure for the probability models and finally realize a high-performance Lepton hardware encoder with low cost and low power consumption.

III. Related Works

A. Hierarchical memory architecture

In hardware design, memory architecture is often a key factor in circuit performance, cost, and power consumption [14] [15] [16]. Hierarchy memory architecture is commonly used to optimize storage costs and access latency [17]. In hierarchy memory architecture, the most frequently used data is placed in a smaller, high-cost, faster cache, and only when the cache addressing fails, that is, cache misses, will the large-capacity, low-cost, slower memory be accessed. In this way, a cheap large-capacity storage structure that exhibits low latency in most cases is realized.

The cache is generally composed of N-way Set-Associative units. If a piece of data could be placed in a set consisting of a limited number of locations in the cache, the cache is called Set-Associative. If the set contains N memory spaces, it is called an N-way Set-Associative unit [18]. The N-way Set-Associative unit improves memory utilization by sharing physical memories between input indexes. An example of N-way Set-Associative units is shown in Fig. 3.

Since Lepton's probability model access does not show a temporal or spatial correlation that crucial to the hierarchy memory architecture, the resulting cache misses will severely degrade the throughput of the Lepton hardware encoder. Besides, designing a hierarchical memory architecture for the Lepton hardware encoder also significantly increases its costs.

TABLE III
THE PROBABILITY MODELS DEFINED BY LEPTON

| probability model | Number of bins | probability model | Number of bins | probability model | Number of bins | probability model | Number of bins |
|---|---|---|---|---|---|---|---|
| exp_7x7_0 | 10780 | exp_edge_0 | 2156 | exp_dc_0 | 204 | res_thres_0 | 4096 |
| exp_7x7_1 | 10780 | exp_edge_1 | 2156 | exp_dc_1 | 204 | res_thres_1 | 8192 |
| exp_7x7_2 | 10780 | exp_edge_2 | 2156 | exp_dc_2 | 204 | res_thres_2 | 16384 |
| exp_7x7_3 | 10780 | exp_edge_3 | 2156 | exp_dc_3 | 204 | res_thres_3 | 32768 |
| exp_7x7_4 | 10780 | exp_edge_4 | 2156 | exp_dc_4 | 204 | res_thres_4 | 65536 |
| exp_7x7_5 | 10780 | exp_edge_5 | 2156 | exp_dc_5 | 204 | res_thres_5 | 131072 |
| exp_7x7_6 | 10780 | exp_edge_6 | 2156 | exp_dc_6 | 204 | res_thres_6 | 262144 |
| exp_7x7_7 | 10780 | exp_edge_7 | 2156 | exp_dc_7 | 204 | res_thres_7 | 4096 |
| exp_7x7_8 | 10780 | exp_edge_8 | 2156 | exp_dc_8 | 204 | res_edge_2 | 196 |
| exp_7x7_9 | 10780 | exp_edge_9 | 2156 | exp_dc_9 | 204 | res_edge_1 | 196 |
| exp_7x7_10 | 10780 | exp_edge_10 | 2156 | exp_dc_10 | 204 | res_edge_0 | 196 |
| res_7x7_0 | 1260 | res_7x7_1 | 1260 | res_7x7_2 | 1260 | res_7x7_3 | 1260 |
| res_7x7_4 | 1260 | res_7x7_5 | 1260 | res_7x7_6 | 1260 | res_7x7_7 | 1260 |
| res_7x7_8 | 1260 | res_7x7_9 | 1260 | res_dc_0 | 12 | res_dc_1 | 12 |
| res_dc_2 | 12 | res_dc_3 | 12 | res_dc_4 | 12 | res_dc_5 | 12 |
| res_dc_6 | 12 | res_dc_7 | 12 | res_dc_8 | 12 | res_dc_9 | 12 |
| nz_7x7_0 | 500 | nz_7x7_1 | 260 | nz_7x7_2 | 140 | nz_7x7_3 | 80 |
| nz_7x7_4 | 40 | nz_7x7_5 | 20 | nz_edgex_0 | 512 | nz_edgex_1 | 256 |
| nz_edgex_2 | 128 | nz_edgey_0 | 512 | nz_edgey_1 | 256 | nz_edgey_2 | 128 |
| sign | 66 | | | | | | |

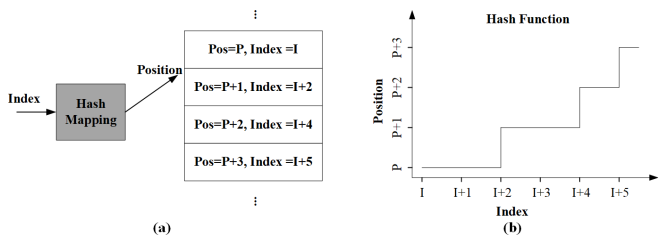

Fig. 4. Example of an ideal hash mapping. (a) shows an ideal read-only database. The data is accessed through the index, and Pos represents the position where the data is stored. Ideally, the data is stored intensively, causing the indexes to be discontinuous. Hash mapping is used to generate the physical storage position based on access index. (b) shows the ideal hash function. It can be seen that the position only increases at the existing index since the non-existent index does not occupy any memory space.

B. Hash mapping

In software, sparse memory access is generally implemented through hash mapping. Hash mapping maps the sparse access index to a compact physical memory space to reduce the required storage capacity. In hash mapping, the index is called Key, the memory address is called Value, and the Key-Value correspondence is determined by the hash function. Hash function has been widely used in data encryption and data management for a long time [19] [20] [21]. When different keys are mapped to the same hash index, a hash collision occurs. In practice, solutions such as linked lists and hash remapping are used to deal with hash collision [22] [23] [24]. The state-of-the-art hash mapping is the learned index structure [11], which is committed to using neural networks to learn a cumulative distribution function-based hash function for large-scale read-only databases. Fig. 4 shows an example of an ideal hash mapping for the read-only database. For read-only databases, it is clear whether an index exists, so the probability distribution function of the index, *P(index)*, and the hash function, *H(index)*, can be represented as

$$P(index_i) = \begin{cases} 1, & if\ index_i\ is\ exiest \\ 0, & otherwise \end{cases} . \quad (3)$$

$$H(index_i) = \sum_{j=0}^{i} P(index_j) \quad . \quad (4)$$

From a statistical point of view, *H(index)* is the cumulative distribution function (CDF) of the index.

Unlike the situation in Kraska et al. [11], the probability model of the Lepton encoder is a write-heavy database, and we must determine the hash function before the binary arithmetic coding starts. Although Ding et al. [25], Hadian et al. [26] proposed several updatable learning-based index structures to improve the access efficiency of large-scale writeable databases, these methods are difficult to be used in hardware design due to the need for real-time expansion of data nodes and retraining of neural network models.

IV. MEMORY OPTIMIZATION FOR PROBABILITY MODELS

In this section, we first analyze the statistical characteristics of probability model access, then introduce an efficient access probability-based hash function for Lepton's probability models, and finally propose a hardware-friendly memory optimization method for the Lepton hardware encoder.

A. Statistical characteristics of probability model access

Once a bin of the probability model is used when encoding an image, we must allocate memory space for it and maintain the memory space until the encoding is completed. The utilization rate of a probability model is defined as the ratio of the number of the used bins to the total number of bins. The test data shows that the utilization rate of the huge probability models declared by Lepton coding algorithm is very low. In addition, there are huge differences between the access probabilities of different bins in the same probability model. For example, for the index of exp_7x7_data, the probability of $\{num\_nz\_left, idx_{zigzag}\} = \{9,17\}$ is very small, since it requires all high-frequency coefficients in current 8x8 block to be non-zero. Similarly, when $idx_{zigzag} = 63$, $num\_nz\_left$ will never be greater than 1. Therefore, we try to provide more memory space for bins with higher access probability, share memory space between bins with lower access probability, and don't allocate memory space for bins that cannot be accessed, thereby greatly reducing the cost of the lepton hardware encoder.

### B. The access probability-based hash function

Inspired by Kraska et al. [11], this work attempts to develop an efficient access probability-based hash function for Lepton. Since Lepton's probability model is a write-heavy database, which requires the hash function to be determined before the data access starts, we cannot establish the corresponding cumulative distribution function based on the actual data distribution. To solve this problem, this work first establishes the probability distribution function of the access index, $P(index_i)$, for each probability model by analyzing the Lepton encoding process of a large number of images, then obtains the cumulative distribution function, $CDF(index_i)$, through partial summation, and finally constructs a similar CDF-based hash function

$$H(index_i) = CDF(index_i) * Mem\_depth. \quad (5)$$

where $Mem\_depth$ is the depth of the physical memory budget. According to the essence of hash mapping and cumulative distribution function, it can be concluded that the memory space occupied by $index_i$ is

$$Space(index_i) = H(index_i) - H(index_{i-1})$$
$$= P(index_i) * Mem\_depth \quad (6)$$

Equation (6) shows that the core idea of $H(index_i)$ is to use the access probability as a weight to allocate memory for each bin. For bins with a small access probability, the memory space allocated to them will be a decimal, which means that they share physical memory space with adjacent bins. Since the probability model must be initialized before encoding every single picture, the validity period of the probability model is limited. In this case, we believe that the probability of hash collision caused by memory sharing between bins with a small access probability is also limited. This hash function concisely achieves the purpose of compressing the required physical storage space with as few hash collisions as possible.

It should be noted that the weight of memory allocation shown in (6) needs simple correction. In practice, one single bin of the probability model only needs one memory space at most, so the maximum weight should not be greater than $1/Mem\_depth$. The final memory allocation weight function $W(index)$ and hash function $H(index)$ are obtained by the pseudo-code shown in Fig. 5. To ensure the normalization of the allocation weights, the clip operation is performed iteratively. Since the hash function only needs to be established once, the iterative operation does not introduce computational costs in the Lepton coding. Another advantage of our method is that since the $\sum W(index)$ will be less than 1 after the clip operation, the renormalization will adaptively amplify the allocation weight for the bins with lower access probability. Therefore, the resulting hash function elegantly removes the dummy bins, and always efficiently allocate memory resource for active bins according to the access probability and the physical memory budget. When the physical memory budget is greater than or equal to the number of active indexes, the generated hash function can theoretically guarantee that no hash

```
1   W(index) = P(index)
2   While(Max(W(index)) > 1/Mem_depth):
3       #clip operation
4       Foreach index_i:
5           if( W(index_i) > 1/Mem_depth):
6               W(index_i) = 1/Mem_depth
7       #renormalization
8       Foreach index_i:
9           W_update(index_i) = W(index_i)/∑W(index)
10      W(index) = W_update(index)
11  Foreach index_i:
12      CDF(index_i) = ∑_{j=0}^{i} W(index_j)
13      H(index_i) = CDF(index_i) * Mem_depth
```

Fig. 5. the pseudo code for establishing the memory allocation weight function $W(index)$ and hash function $H(index)$ according to the access probability distribution function $P(index)$.

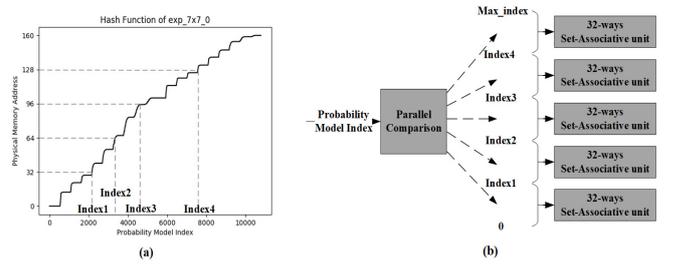

Fig. 6. Example of the proposed hardware-friendly memory optimization method. (a) shows the access probability-based hash function of the probability model exp_7x7_0. The 10780 indexes of exp_7x7_0 are mapped to 160 physical addresses by the hash function, and then the entire index range is divided into five intervals according to the physical address, each interval shares 32 memories (M=5, N=32, Mem_depth=160). (b) shows the access of the probability model in the Lepton hardware encoder. When the probability model exp_7x7_0 is accessed, the hardware encoder determines the interval in which the index is located through parallel comparison, and then enables the corresponding 32-ways Set-Associative unit to complete the read and write operations.

collision occurs. Since the memory access of most applications has distinct statistical characteristics, we argue that other sparse memory accesses can also benefit from this hash function.

### C. The hardware-friendly memory optimization method

After establishing the hash function, we need to express it in the application. These functions are often difficult to express as simple analytical forms. In fact, the main purpose of the learned index [11] [25] [26] is to fit the CDF by using the neural network. To utilize the proposed hash function in hardware design, we further propose a hardware-friendly memory optimization method based on the N-way Set-Associative units. The proposed method divides the index range into M intervals according to the access probability-based hash function, and then stores the M-1 boundary indexes. When the probability model is accessed, we first identify the interval in which the index is located through parallel comparison and then determine the corresponding N-way Set-Associative unit. The indexes in one interval share the N physical memory spaces in an N-way Set-Associative unit. Obviously,

$$M = Mem\_depth/N. \quad (7)$$

The N-way Set-Associative unit sequentially allocates memory for the input index and records the allocation status at the same time. For the input index, the N-way Set-Associative unit first queries whether it already has a memory space, if so, the Set-Associative unit directly gets the memory address, otherwise, it allocates memory for this index. When M=Mem_depth, this scheme is completely equivalent to the proposed hash function. When M=1, this scheme is completely equivalent to the fully-associative scheme [7]. The memory utilization and the complexity of Set-Associative unit will both increase as M decreases. If more than N different indexes are accessed in any index interval during encoding an image, the probability model overflow occurs. Fig. 6 is an example of the proposed memory optimization method.

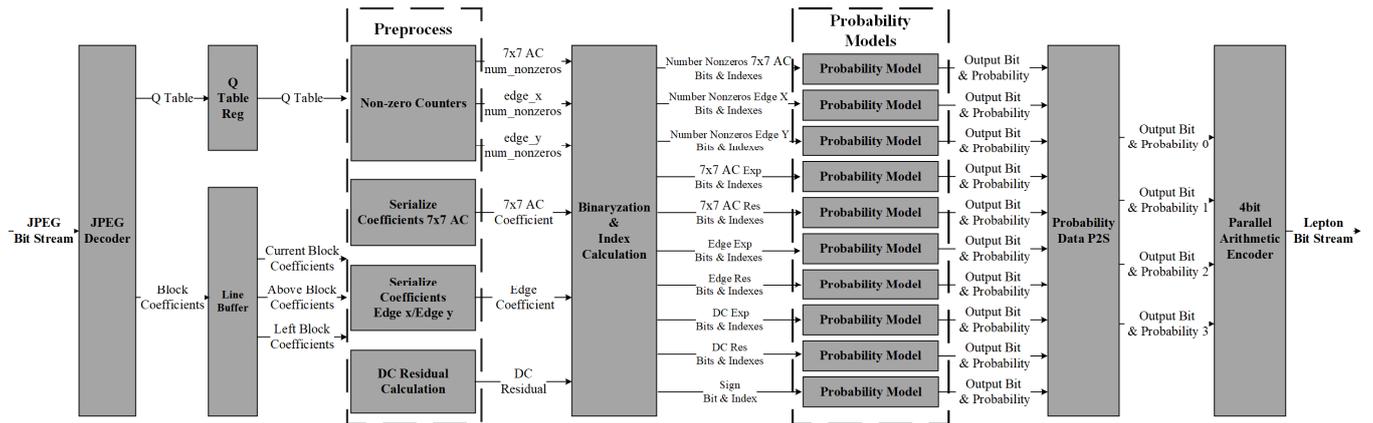

Fig. 7. The structure of Lepton hardware encoder.

Compared with using the hash function alone, the proposed memory optimization method has the following advantages: 1. It only needs to store M-1 boundary indexes to realize the purpose of allocating memory resources according to the access probability, avoiding fitting the complex hash functions. 2. Since different indexes share N physical memories, this method has a higher memory utilization. 3. This method is meaningful for both software and hardware design. For the software design, we can use B-trees to store and index these boundary indexes and use linked lists to implement N-way Set-Associative units, which makes it possible to implement the access probability based-hash function through a simple traditional index structure. For the hardware designs, this method transforms the complex function fitting problem into a series of hardware-friendly parallel comparison operations, which greatly reduces the design complexity and implementation cost.

### D. probability model overflow solution

Although the proposed memory optimization method can ensure that no probability model overflow occurs when the memory capacity is greater than the number of active indexes, roughly expanding the memory capacity may not be the optimal choice. When the overflow rate is low, the marginal benefit of expanding the memory capacity will gradually decrease. To further reduce the cost of the Lepton hardware encoder without significantly degrading the processing performance, this work adopts a combination of software and hardware designs. When the probability model overflow occurs, the Lepton hardware encoder generates an interrupt signal and records the overflow index, and then the interrupt service program will call the software lepton encoder to re-encode the current image. As long as the overflow rate is low enough, the impact of such operations on the overall throughput is negligible. In addition, the interrupt service program can also analyze the overflow indexes, refresh the boundary indexes according to the updated access probability-based hash function, and adaptively reduce the overflow rate.

## V. VLSI ARCHITECTURE

### A. Lepton hardware encoder

The structure of the Lepton hardware encoder is shown in Fig. 7. The JPEG images are first decoded into quantized DCT coefficients and quantization table data. The Line Buffer module saves the quantized coefficients and outputs the current block, the left reference block, and the upper left reference block to the subsequent module. The preprocess part calculates the number of non-zero coefficients in the 7x7 AC, x edge, and y edge regions of the current block, the residual value of the DC coefficient, and serially outputs the coefficient data that needs to be coded in the 7x7 AC, x edge and y edge regions. The binarization and index calculation module converts the coefficient data into exponent, sign, and residual by using the Exp-Golomb code, and generates the probability model index for each bit. The probability model reads and updates the corresponding bin according to the input index, calculates the probability value. The probability data p2s module receives the coding bits and the probability values from the probability models in parallel and outputs them to the 4-bits parallel arithmetic encoder in the order specified by Lepton to generate the final Lepton code stream.

The probability model part integrates all the probability models defined in the Lepton coding algorithm, which results in the main cost of the Lepton hardware encoder. Fig. 7 only shows the schematic diagram of the probability models according to the data category. In fact, the hardware encoder contains 77 independent instances of probability models, corresponding to different memory capacities. Therefore, the cost, performance, and reusability of the probability model hardware structure are carefully considered.

### B. probability models

As shown in Fig. 8, the probability model is mainly composed of enable generator, N-way Set-Associative unit array, and probability model controller. The enable generator compares the input index with the M-1 boundary indexes and then generates the enable signal for the corresponding N-way Set-Associative unit through simple logical operations. The i-th N-way Set-Associative unit includes N 1-bit initialization flags and N $R_i$-bit index records, $R_i$ is determined by

$$R_i = Roundup(\log_2(BI_{i+1} - BI_i)). \quad (8)$$

where $BI_i$ is the i-th boundary index. When the input index is valid, the N-way Set-Associative unit will compare the input index with the R-bit index records. If there is a matching index record, we then directly get the corresponding memory address, otherwise, it is further judged whether there is free memory space. If the memory space is not full, the N-way Set-Associative unit will allocate memory to the input index, set the corresponding initialization flag, and then add the input index into the index records, otherwise the circuit will raise the overflow signal and report the current index.

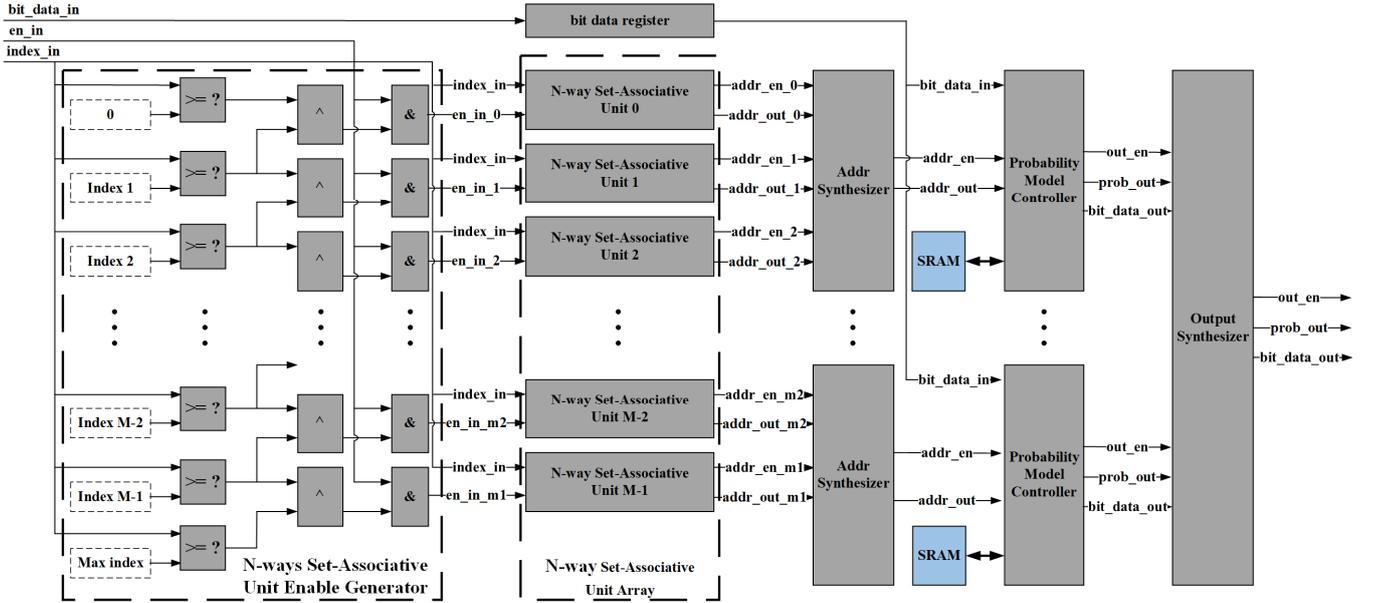

Fig. 8. The structure of the probability models.

TABLE IV
THE UTILIZATION RATES OF EXP_7X7_0-EXP_7X7_10

| probability model | exp_7x7_0 | exp_7x7_1 | exp_7x7_2 | exp_7x7_3 | exp_7x7_4 | exp_7x7_5 |
|---|---|---|---|---|---|---|
| utilization rate | 50.78% | 49.59% | 46.42% | 43.22% | 40.03% | 33.75% |
| probability model | exp_7x7_6 | exp_7x7_7 | exp_7x7_8 | exp_7x7_9 | exp_7x7_10 | |
| utilization rate | 24.25% | 14.73% | 8.07% | 3.17% | 0.24% | |

The probability model controller reads and updates the probability model data according to the memory address and enable signal, and calculates the corresponding probability value based on the read data.

It is worth noting that, for the largescale memory management circuits, the quantity, shape, and coupling relationship of the memories often pose severe challenges to the layout and routing in the back-end process [27]. In our design, the address synthesizer allows the physical memory depth managed by each probability model controller to be flexibly adjusted by a step of N. If the address synthesizer synthesizes the outputs of K N-way Set-Associative units, the subsequent probability model controller will manage a physical memory depth of K × N. The final output is generated by the output synthesizer according to the enable signal.

In summary, the advantages of the proposed VLSI structure are: 1. As shown in Fig. 8, the comparison operation is executed in parallel and the subsequent output synthesizer can be easily designed as a pipeline, the circuit delay does not rise as M increases. It means that the proposed structure can efficiently adapt to the large-scale probability model; 2. The circuit structure is very simple, and the M, N, R, and K mentioned above are all configurable parameters, which makes the modules highly reusable. 3. The memory shape can be selected flexibly, which is conducive to the place and routing in the back-end process. 4. The boundary indexes can be updated through the configurable register, allowing the interrupt service program to dynamically refresh the hash function.

VI. EXPERIMENTS AND RESULTS

*A. Image set and statistical method*

Since the interrupt service program handles the probability model overflow in the unit of the entire image, we use a similar standard to calculate the overflow rate of the Lepton hardware encoder, that is, as long as any probability model overflow occurs, the whole encoding of the current image is considered to be a failure, and the overflow rate is the proportion of images that have failed to encode. Under this regulation, the higher the image resolution, the lower the tolerance for probability model overflow. In order to get more convincing results, we establish experiment based on the high-resolution image sets DIV2K and Kaggle4K. DIV2K is a high-resolution image set of the NTIRE 2017 challenge on single image super resolution, including 800 training images, 100 validation images, and 100 test images [28]. Kaggle is an open-source platform for developers and data scientists to host machine learning competitions and share databases. Kaggle4K is a high-resolution image set on Kaggle, containing 2056 4K images [29].

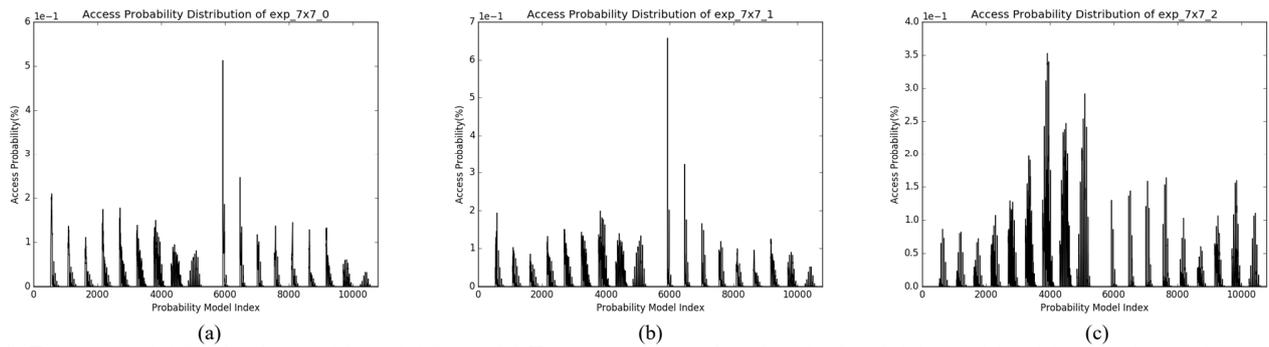

Fig. 9. The access probability distribution of the probability model. The horizontal axis is the index of each probability model, and the vertical axis is the probability that the index is accessed during the encoding process. (a), (b), (c) correspond to the probability models exp_7x7_0, exp_7x7_1, exp_7x7_2, respectively.

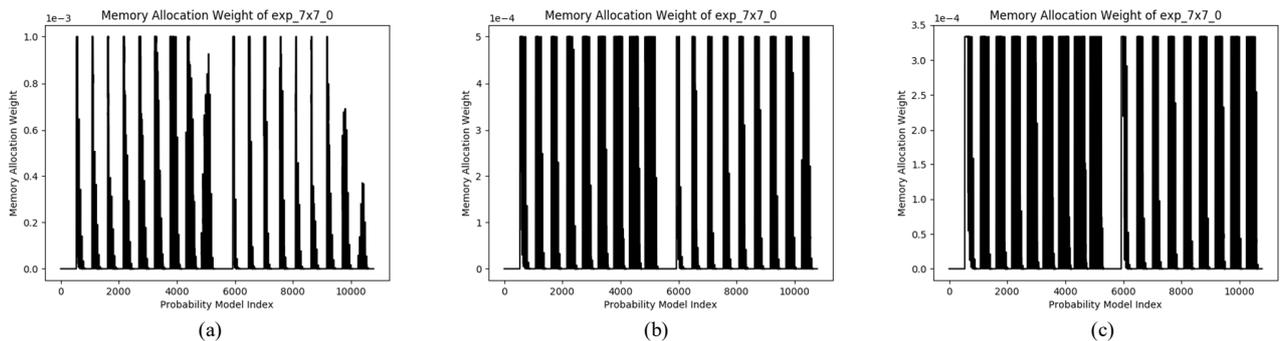

Fig. 10. The memory allocation weight function of exp_7x7_0 under different mem depths, the horizontal axis is the index of the probability model, and the vertical axis is the memory allocation weight of each index, (a), (b), (c) correspond to the memory depth of 1000, 2000, 3000, respectively.

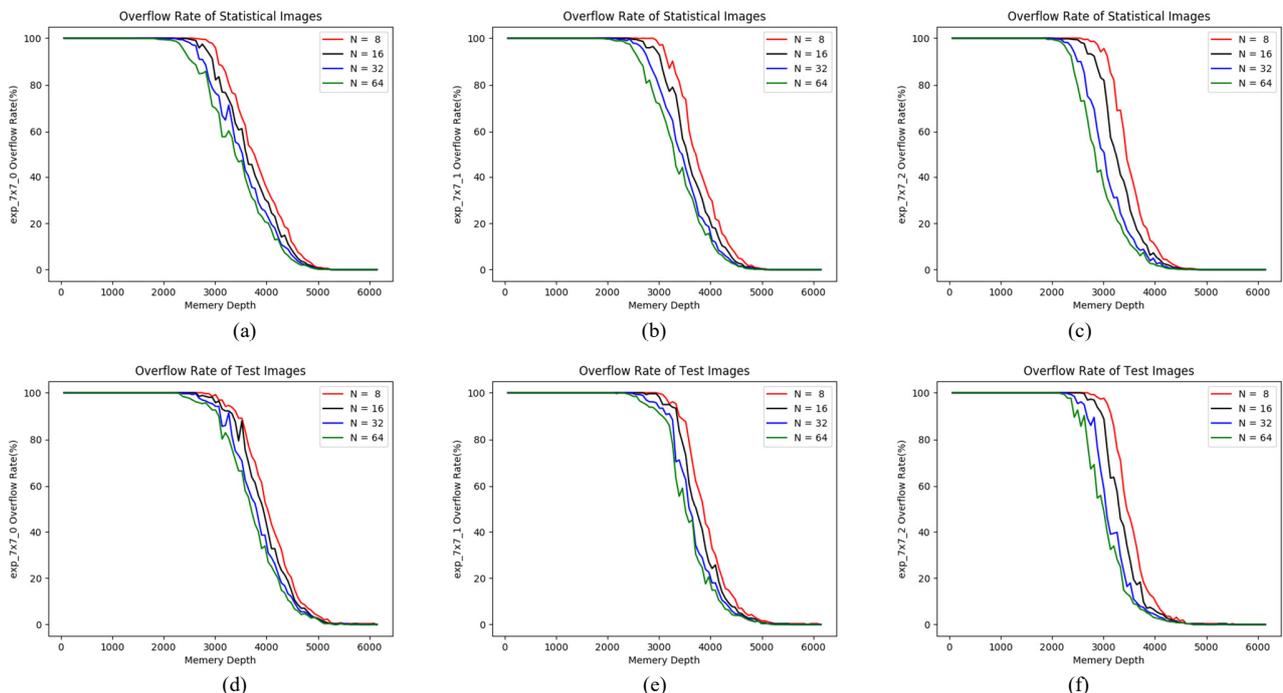

Fig. 11. The overflow rate of the probability models exp_7x7_0-exp_7x7_2 with different memory depths. (a), (b), (c) correspond to the overflow rate of exp_7x7_0-exp_7x7_2 on the statistical images, (d), (e), (f) correspond to the overflow rate of exp_7x7_0-exp_7x7_2 on the test images, respectively.

We use the Lepton reference model [13] to analyze the statistical characteristic of the probability model access. In order to demonstrate the generalization ability of the proposed memory optimization method, we extract statistical features on Kaggle4K 0001-2000 and DIV2K 0001-0799, and calculate the overflow rate on Kaggle4K 2001-2056 and DIV2K 0800-1000. It can be seen from Table III that exp_7x7_0-exp_7x7_10, res_7x7_0-res_7x7_9, exp_edge_0-exp_edge_10, and res_thres_0-res_thres_7 are the main part of Lepton's probability models, so the memory optimization method is evaluated on these probability models. For brevity, we take probability models exp_7x7_0-exp_7x7_10 as examples in the discussion and just show the optimization results of the other probability models.

TABLE V
THE MEMORY DEPTH REQUIRED BY EACH PROBABILITY MODELS

| Probability Models | N = 8 | N = 16 | N = 32 | N = 64 |
|---|---|---|---|---|
| exp_7x7_0-exp_7x7_10 | 36544 | 35520 | 33152 | 32128 |
| exp_edge_0-exp_edge_10 | 10624 | 10112 | 9536 | 9472 |
| res_7x7_0-res_7x7_9 | 4160 | 4096 | 3968 | 3968 |
| res_thres_0-res_thres_7 | 15872 | 15552 | 14336 | 13760 |
| Total of Lepton | 73050 | 71130 | 66842 | 65178 |
| Memory Saving (%) | 89.34 | 89.62 | 90.24 | 90.49 |

TABLE VI
THE AREA OF THE PROBABILITY MODELS

| Probability Models | The original method | The optimized method (N=32) |
|---|---|---|
| exp_7x7_0-exp_7x7_10 | 5.30 | 3.59 |
| exp_edge_0-exp_edge_10 | 1.06 | 0.90 |
| res_7x7_0-res_7x7_9 | 0.56 | 0.35 |
| res_thres_0-res_thres_7 | 21.70 | 2.07 |
| Total of Lepton | 30.59 | 8.88 |

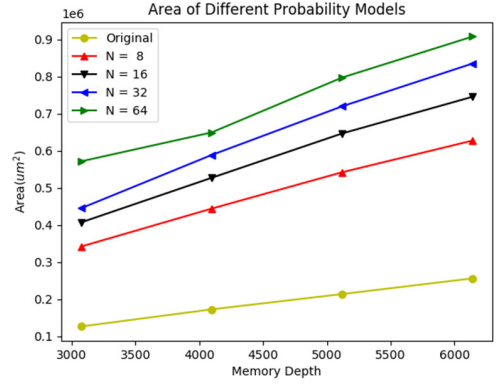

Fig. 12. The area of different probability models. We use SIMC 65nm process to synthesize the proposed probability models for the cases of N = [4,8,16,32] and Memory Depth = [3072,4096,5120,6144]. The clock frequency is 800MHz and no timing violation exists.

*B. Results of memory optimization*

The utilization rates of exp_7x7_0-exp_7x7_10 are shown in Table IV. It can be seen that the utilization rates of each probability model in Lepton are very different, and most of them are quite low, which implies that Lepton's probability models have an obvious necessity of optimization.

The access probabilities of different indexes of the probability model exp_7x7_0-exp_7x7_2 are shown in Fig. 9. We can see that there are indexes with higher access probability, indexes with lower access probability, and indexes that cannot be accessed in each probability model.

Fig. 10 shows the memory allocation weight function of exp_7x7_0 when memory depth is 1000, 2000, 3000 respectively. It can be seen that as the memory depth increases, the proposed memory optimization method not only accurately clips the upper limit of the allocation weights but also adaptively amplifies the allocation weights of the indexes with lower access probability.

Fig. 11 shows the relationship between the overflow rate and the memory depth of probability model exp_7x7_0-exp_7x7_2 when N is 8, 16, 32, and 64 respectively. It can be seen that the proposed memory optimization method can not only perfectly fit the statistical images, but also perform well on the test images.

Based on the criteria that no overflow occurs on both statistical images and test images, we calculate the required physical memory depth of probability model exp_7x7_0- exp_7x7_10, res_7x7_0-res_7x7_9, exp_edge_0-exp_edge_10, and res_thres_0-res_thres_7, and show the results in Table V. It means that if a software solution based on B-tree and linked list is adopted to implement the proposed memory optimization method, the memory space required by the Lepton probability models can be reduced by 89.34%~90.49% at the cost of slightly increasing the access delay.

*C. The area of the probability models*

In order to illustrate the area reduction of the proposed probability models, this work synthesizes the circuits with the SIMC 60nm process and uses the original scheme which allocates dedicated memory for each index as a comparison. The target clock frequency is set to 800MHz and no timing violation exists.

The area of the probability models with different memory depths and N values is shown in Fig. 12. Since additional logic resources are required to manage the memories and maintain the allocation status, the area of the proposed probability model is naturally larger than that of the original method when their memory capacities are equal. In addition, it can be seen from Fig. 12 that for each value of N, the area of the probability models and the memory depth are approximately in a linear relationship. Therefore, we use the Mean Area (MA) of the memories in each probability model to evaluate the area benefit achieved by the proposed memory optimization method.

For the original scheme, the required memory depth equals to the index range, so its total area is given by

$$Area_{orig} = MA_{orig} * index\_range. \quad (9)$$

Similarly, the area of the proposed probability model is

$$Area_{opt} = MA_{opt} * Mem\_depth. \quad (10)$$

Therefore, when

$$Mem\_depth/index\_range < MA_{orig}/MA_{opt}, \quad (11)$$

the proposed probability model will reduce the area of the Lepton hardware encoder. For the memory optimization method, we define memory reduction as the ratio of the reduced physical memory to the index range. Taking N=32 as an example, according to Fig. 12, $MA_{orig}/MA_{opt}$ is about 0.3, that is, for the probability models with a memory reduction greater than 70%, the proposed memory optimization scheme should be used. Therefore, we use the proposed memory optimization method to implement the probability models that meet this criterion, while the other probability models are implemented by the original scheme. The area results are shown in Table VI. It can be seen that the proposed memory optimization method reduces the total area of the probability models by 70.97%.

TABLE VII
THE SYNTHESIS RESULT OF LEPTON HARDWARE ENCODER

|  | Area ($mm^2$) | Energy Efficiency (MB/J) | Frequency (MHz) | Throughput (images/Sec) |
|---|---|---|---|---|
| DropBox's | - | 0.03 | - | 5.75 |
| Ours | 10.13 | 146.97 | 617.59 | 317.69 |

*D. The Lepton hardware encoder*

We synthesize the proposed Lepton hardware encoder with the SIMC 60nm process and compare the results with DropBox's software-based solution in Table VII. The energy efficiency is defined as the energy consumed for generating 1MB Lepton bitstream and the throughput is defined as the number of FHD images that can be encoded per second. It is worth noting that although the test results show that the Lepton software model [13] requires an average of 8.9 seconds to encode one FHD image on a single CPU core, which means that the 16-core parallel encoding could process 1.8 images per second, we still quote the data in Horn et al. [5], which is 5.75 images per second, as a comparison. Testing results show that the Lepton hardware encoder can encode more than 317.69 FHD images per second, and no overflow occurs on the test images. Compared to the CPU-based software encoder, the throughput of the Lepton hardware encoder is increased by 55.25 times, and the energy efficiency is increased by more than 4899 times. In addition, the manufacturing cost of Lepton hardware encoders is much lower than the CPU core.

VII. CONCLUSION

In this work, we greatly reduce the memory requirements of Lepton's probability models and implement a high throughput, high energy efficiency, low-cost Lepton hardware encoder. The hash function and hardware-friendly memory optimization method proposed in this work achieve the purpose of effectively allocating memory space according to access probability and memory budget with low design complexity. In addition, the tremendous energy efficiency improvement of the proposed Lepton hardware encoder can greatly reduce the operating costs of data centers. Since actual sparse data access usually has significant statistical characteristics, we believe that our method can benefit a large number of similar data access applications.